\documentclass[twocolumn, showpacs,preprintnumbers,amsmath,amssymb]{revtex4}

\usepackage{graphicx}
\usepackage{dcolumn}
\usepackage{bm}

\begin{document}

\title{Electronic and magnetic influences of a stacking fault in cobalt nanoscale islands on the Ag(111) surface}
\author{Keiji Doi$^1$}
\author{Emi Minamitani$^2$}
\author{Shunji Yamamoto$^1$} 
\author{Ryuichi Arafune$^3$}
\author{Yasuo Yoshida$^{1}$}
\thanks{Corresponding author; yyoshida@issp.u-tokyo.ac.jp} 
\author{Satoshi Watanabe$^2$}
\author{Yukio Hasegawa$^1$}
\affiliation{$^1$The Institute of Solid State Physics, the University of Tokyo, Kashiwa 277-8581, Japan. \\
$^2$Department of Materials Engineering, The University of Tokyo, Tokyo 113-8656, Japan.\\
$^3$International Center for Materials Nanoarchitectonics, National Institute for Materials, Science, 1-1 Namiki, Ibaraki 304-0044, Japan.}
\date{\today}

\begin{abstract}
Utilizing spin-polarized scanning tunneling microscopy and spectroscopy,  
we found coexistence of perpendicularly and in-plane magnetized cobalt nanoscale islands on the Ag(111) surface, and 
the relationship between the moir\'e corrugation amplitude and the magnetization direction of the islands; the islands with the stronger moir\'e corrugation show the perpendicular magnetization, and the ones with the weaker moir\'e corrugation do the in-plane. 
Density functional theory calculations reproduce the relationship 
and explain the differences between the two types of the islands with an fcc stacking fault in the intrinsic hcp stacking of cobalt. 

\end{abstract}

\maketitle
\section{introduction}
Magnetic thin films with perpendicular magnetic anisotropy (PMA) have attracted great intersts in the last decade 
from the viewpoint of technological applications such as a high-density magnetic storage \cite{pma storage}, 
a magnetic random access memory \cite{pma mram} and other spintronics applications \cite{pma mtj}. 
Cobalt (Co), a 3d transition metal, is a typical material for experimental studies of PMA in magnetic nanoscale structures 
because of the advantageous properties; the strong uniaxial magnetic anisotropy and 
the usability to control the properties in various shapes; single atoms \cite{gambadella1, gambadella2, ibm}, nanoscale islands \cite{ossie, focko, jessica, co/au111}, and thin films \cite{co/pd111, co/rh111}. 
There are reports on the emergence of PMA of Co single atoms on Pt(111) \cite{gambadella1, gambadella2} and MgO thin films \cite{ibm}, 
and also on that of Co nanoscale structures formed on noble metals \cite{ossie, focko, jessica, co/au111, co/pd111, co/rh111}. 
Especially, nanosized Co clusters or islands are attractive for ultrahigh density magnetic recording applications \cite{hd ieee, hd science}, 
and therefore, it is important to identify and characterize detrimental factors.  
One of the factors which drastically reduce the magnetic anisotropy 
is stacking faults in hcp Co structures \cite{sf hinata, sf sokalski, sf theory}. 
However, a microscopic investigation on this issue has not be done so far.  

Here we investigate the PMA of hcp Co nanoscale islands on a Ag(111) substrate and 
the effect of a stacking fault on it using a combination of spin-polarized scanning tunneling microscopy (SPSTM) experiments and density functional theory (DFT) calculations. 
Cobalt islands with triangular or hexagonal shapes and thicknesses ranging from 5 to 8 monolayers (MLs) are formed on the substrate. 
They are found to show a moir\'e pattern on them. 
The islands are categorized into two groups with respect to the corrugation amplitude of the pattern in STM topographies taken at -0.2 V; 
the islands with stronger and weaker moir\'e corrugation amplitudes.  
By nanoscale magnetometry based on SPSTM, 
we found that the islands with the stronger moir\'e corrugation show hysteretic magnetization curves with perpendicular magnetizations  
while the ones with the weaker moir\'e corrugation do in-plane magnetizations without hysteresis.   
DFT calculations reveal the difference in the moir\'e corrugation amplitude comes from the difference in the stacking structure; 
hcp stacking for the islands with the stronger moir\'e corrugation amplitude and hcp stacking with an fcc stacking fault, which we call pseudo-hcp hereafter, 
for ones with the weaker moir\'e corrugation amplitude. 
By calculating magnetic anisotropies of the islands, 
we figured out that PMA of the Co nano-islands is significantly reduced due to the stacking fault 
and even becomes weaker than the shape anisotropy, resulting in the variation of the easy magnetization axis. 

\section{experimental details}

All measurements were performed in a low-temperature ultra-high vacuum STM (Unisoku USM-1300S with an RHK controller R9). 
The sample and the tip can be cooled with liquid He, and an external magnetic field can be applied perpendicular to the sample surface. 
For the detection of spin-dependent tunneling current, we used a bulk Cr tip \cite{annika, oka, romming, 1st} that was electrochemically etched from a Cr rod. 
The tip made of the antiferromagnetic material minimizes stray field, which may modify the magnetic properties of samples. 
In our SPSTM measurements, the tip exhibited weak perpendicular magnetization whose direction can be flipped by the external magnetic field. 
This is probably due to the presence of a Co nano cluster picked up on the tip apex. 
The tip magnetization was characterized by taking a hysteresis curve of spin-polarized tunneling conductance in the magnetic field \cite{rodary}. 
For the differential conductance (d$I$/d$V$) spectroscopy, we used a lock-in technique with a modulated sample bias voltage of 842 Hz and 10 m$V_\mathrm{RMS}$. 
A two-dimensional d$I$/d$V$ mapping was taken simultaneously with a topographic image with the modulation voltage of 50 m$V_\mathrm{RMS}$.  

The Co/Ag(111) sample was prepared by firstly cleaning the substrate in a standard method: repetitive Ar sputtering and annealing at $\sim$1000 K, and then depositing $\sim$ 1 ML Co (99.99 \% purity) on it by electron bombardment heating.
The deposition rate was set $\sim$ 0.1 ML/s, and during the deposition the sample was kept at room temperature. 

\section{Theoretical calculation}
DFT calculations were performed by the plane-wave-based Vienna \textit{Ab Initio} Simulation Package (VASP) \cite{1,2} with the projected augmented wave (PAW) method \cite{3}. 
The exchange and correlation were described at the level of generalized gradient approximation (GGA). 
For the exchange-correlation functional, 
we use that determined by Perdew-Burke-Ernzerhof \cite{4}. 
The Co nanoisland on Ag(111) was modeled by the slab model consisting of a Co 5-layer of (7$\times$7) periodicity on a 5-layer Ag(111) substrate of (6$\times$6) periodicity and a vacuum of $\sim 18 \mathrm{\AA}$ thick along the surface normal. 
During the structural optimization, 
the position of atoms in the Co layers and the top 4-layers of the Ag(111) slab were optimized without any constraint until the forces on individual atoms were less than 0.02 eV/ $\mathrm{\AA}$. 
For this system, we set the energy cutoff for the plane wave basis at 400 eV, 
and the Brillouin zone was sampled with a $k$-point only at  $\mathrm{\Gamma}$ point because of the large size of the supercell. 

In the magnetic anisotropy energy (MAE) calculation of the free-standing 5 atomic layer of Co with (1$\times$1) periodicity, 
we set the cutoff energy at 600 eV and use 31$\times$31$\times$1 Monkhorset-Pack $k$-point mesh \cite{5}. 
We checked the convergence of the MAE on the cutoff energy and the $k$-point sampling, and found that the above setting is enough. 
The MAE is estimated from the difference between the total energy for the c-axis (out-of-plane) magnetization and the a-axis (in-plane) magnetization.

\section{Experimental results}

\begin{figure}
\includegraphics[width=1\columnwidth,keepaspectratio]{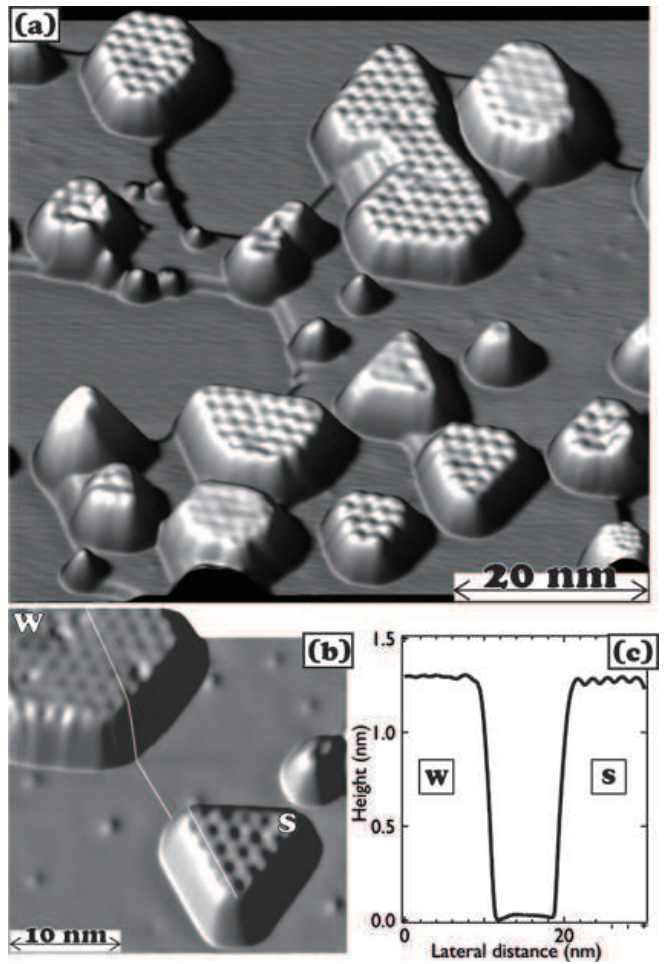}
\caption{\label{fig:fig1}
(a) Overview image of Co-deposited Ag(111) surface (sample bias voltage $V_\mathrm{S}$= -0.2 V, tunneling current $I_\mathrm{T}$= 1 nA).
Co islands with thicknesses ranging from 5 to 7 MLs are observed. 
(b) Constant-current image of the 6 ML Co islands on Ag(111) ($V_\mathrm{S}$= -0.2 V, $I_\mathrm{T}$= 0.5 nA). 
The island types are labeled with S (strong moir\'e) and W (weak moir\'e). 
(c) Cross-sectional profile as indicated in (b), along the two islands that has the same thickness. S and W indicate the island types.  
}
\end{figure}

\begin{figure}
\includegraphics[width=1\columnwidth,keepaspectratio]{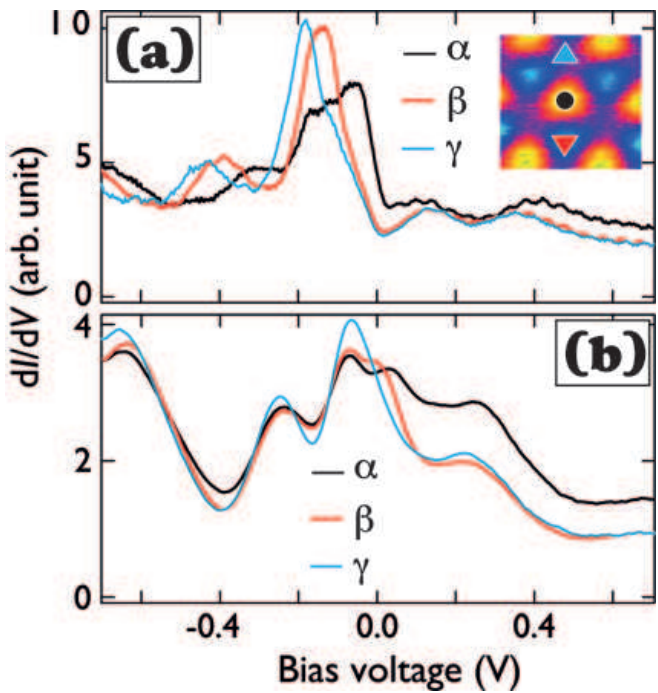}
\caption{\label{fig:fig2}
(Color online) 
Spin-averaged d$I$/d$V$ spectra taken at three different sites ($\alpha$, $\beta$, $\gamma$) in the moir\'e pattern on the strong (a) and weak (b) moir\'e islands. 
The tip-sample separation was stabilized with the condition of 
$I_\mathrm{T}$= 1.0 nA and $V_\mathrm{S}$=-0.7 V.  
The inset of (a) is an STM image (3.1 nm x 3.1 nm, $I_\mathrm{T}$= 1.0 nA and $V_\mathrm{S}$=-0.2 V) showing exact positions of $\alpha$ (black circle), $\beta$ (red triangle), and $\gamma$ (blue triangle) sites.
}
\end{figure}

Figure 1(a) shows a typical STM image taken on the Co-deposited Ag(111) surface 
at 5 K and in ultra-high vacuum conditions. 
Co islands with three-fold symmetric shapes and thicknesses ranging from 5 to 7 MLs are observed. 
Since cobalt has the hcp structure intrinsically, 
these shapes suggest that the island surface has a close-packed plane and the island grows to the (0001) orientation. 
One can also easily notice a corrugation on the surfaces 
whose orientation is along the triangular or hexagonal edge of the islands 
which corresponds to the atomic row direction.   
The periodicity of the corrugation is 1.74 nm agreeing with the previous report on 3 ML Co islands on Ag(111) \cite{berndt}. 
The pattern is explained as a moir\'e pattern coming from a lattice mismatch between the Co overlayer and the Ag(111) substrate \cite{berndt}, 
whose interatomic distance is 0.251 nm and 0.289 nm, respectively. 
In the pattern, eight Co atoms are situated on seven Ag atoms (7 atomic spacing of Co matches to 6 atomic spacing of Ag) and 
the average interatomic distance of Co in the moir\'e is estimated as 0.248 nm, smaller than the bulk \cite{berndt}. 
Unlike the case of double-layer Co on Cu(111) \cite{kern, ossie2, okascience}, the periodicity does not change with the bias voltage, ruling out the possibility that the corrugation is due to 
surface standing waves. 

The most interesting observation here is that islands have different moir\'e corrugation amplitudes at the sample bias voltage of -0.2 V; 
some of them have larger moir\'e corrugation than the others. 
We observed both upward and downward triangular islands, 
implying their different stacking at the interface between the islands and the substrate \cite{ossie}.  
But it is clear from the image that this interfacial stacking difference does not affect the moir\'e corrugation amplitude since 
some of the triangular islands point to the same direction but having different moir\'e corrugation amplitudes. 
Therefore the interfacial stacking difference is not the origin of the moir\'e difference. 
To compare the difference in the moir\'e corrugation amplitude, 
we focus on two Co islands with 6 ML thinckness shown in Fig. 1(b). 
The right island in Fig. 1(b) has larger moir\'e corrugation amplitude (height modulation: 25 pm) than the left one (8 pm), as clearly demonstrated in a cross-sectional profile of Fig. 1(c). 

In order to clarify the origin of the corrugation difference, we have investigated electronic structures of 
two types of islands by spin-averaged d$I$/d$V$ spectroscopy at three different sites in the moir\'e 
as shown in Figs. 2(a) and (b). 
The protrusions and two hollow sites in the moir\'e structure, that correspond to the top, hcp, and fcc sites of Ag(111), 
are referred as $\alpha$, $\beta$, and $\gamma$ sites, respectively \cite{berndt}. 
In the spectra taken on islands with the stronger moir\'e corrugation amplitude (Fig. 2(a)), 
two distinct peaks are observed below the Fermi energy ($E_\mathrm{F}$) for all sites. 
The peak positions vary significantly depending on the sites. 
At $\alpha$ site, a large peak appears around -0.1 V with a shoulder around -0.2 V followed by a small peak at -0.3 V. 
At $\beta$ ($\gamma$) site, the large peak appears sharper than that at $\alpha$ site with the energy at -0.15 V (-0.2 V) and 
the small peak does at -0.4 V (-0.45 V). 
In contrast, on islands with the weaker moir\'e corrugation amplitude, 
these peaks are shifted to -0.08 V and -0.25 V and appear almost at the same energies for all sites. 
Since the contrast in constant-current STM images depends on the integral of the local density of states (LDOS) between $E_\mathrm{F}$ and the set bias voltage, 
the variations in the peak positions around the voltage affects the STM contrast. 
We thus conclude that the difference in the moir\'e corrugation amplitude between the islands observed in STM topographies at  V$_\mathrm{S}=$-0.2 V is due to the difference in the electronic structures. 
(Hereafter in this paper, we call the islands strong- or weak-moir\'e islands 
with respect to the moir\'e corrugation amplitudes in STM topographies at V$_\mathrm{S}=$-0.2 V.)
These kinds of distinct peaks below the $E_\mathrm{F}$ originate from the minority- and majority-spin 
$d$-bands in the case of Co/Cu(111), 
which was confirmed by comparing the d$I$/d$V$ spectra with calculated LDOS \cite{kern, rastei}. 
The similar assignment are also done for the current system by Gopakumar {\it et al.}; 
-0.2 V (-0.4 V) peak of the strong moir\'e island is ascribed to the minority (majority) spin state \cite{berndt}.  

\begin{figure*}[htbp]
\includegraphics[width=2.05\columnwidth,keepaspectratio]{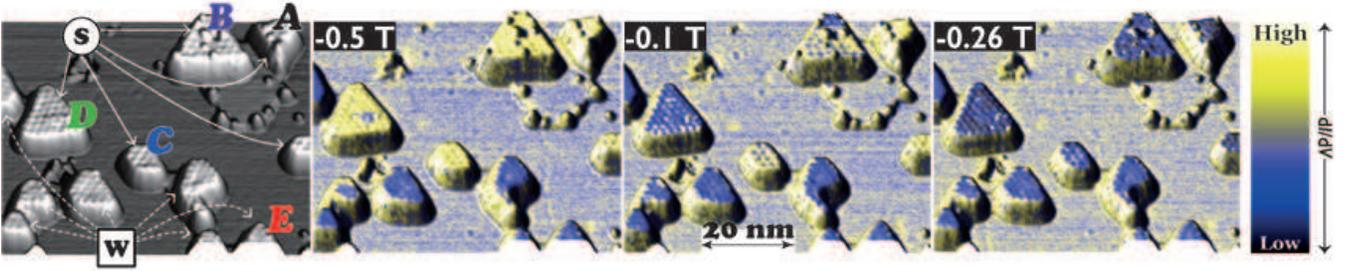}
\caption{\label{fig:fig3}
(Color online) Topography (leftmost) and spin-resolved d$I$/d$V$ colorized topographies 
at magnetic fields of -0.5 T, -0.1 T, and -0.26 T (in the downward field sweep) ($V_\mathrm{S}$= -0.4 V, $I_\mathrm{T}$= 1 nA). 
More detailed field dependence of the d$I$/d$V$ mapping is available as a movie \cite{supplement}.  
The island types are labeled S (strong moir\'e) and W (weak moir\'e) in the topography. 
Thickness of the islands; A: 7 ML, B: 8 ML, C: 6 ML, D: 6 ML, E: 6 ML. }
\end{figure*}

\begin{figure}[htbp]
\begin{center}
\includegraphics[width=0.97\columnwidth,keepaspectratio]{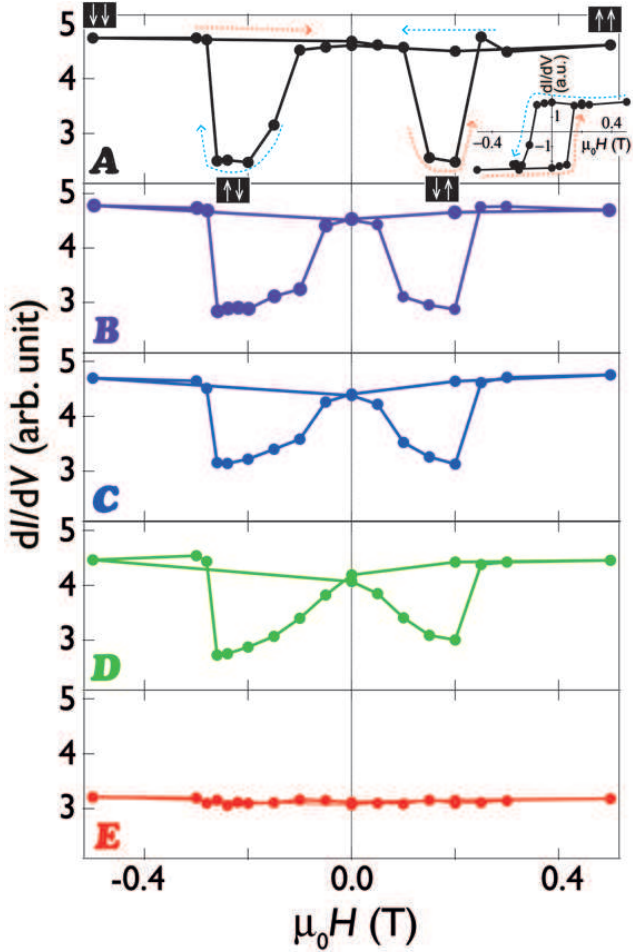}
\caption{\label{fig:fig4}
(Color online) Magnetic-field dependences of the d$I$/d$V$ extracted from the islands A to E in the spin-resolved d$I$/d$V$ images shown in Fig. 3 
($V_\mathrm{S}$= -0.4 V, $I_\mathrm{T}$= 1 nA). 
The magnetic field is applied perpendicular to the surface. 
A butterfly-shaped curve was obtained on the strong moir\'e islands A to D but not on the weak moir\'e island E. 
The inset of the top panel is the hysteretic d$I$/d$V$ curve of 
the island A estimated from the butterfly-shaped curve given the 
flippings of the tip magnetization direction at -0.28 T and 0.25 T. }
\end{center}
\end{figure}

\begin{figure}[htbp]
\begin{center}
\includegraphics[width=1\columnwidth,keepaspectratio]{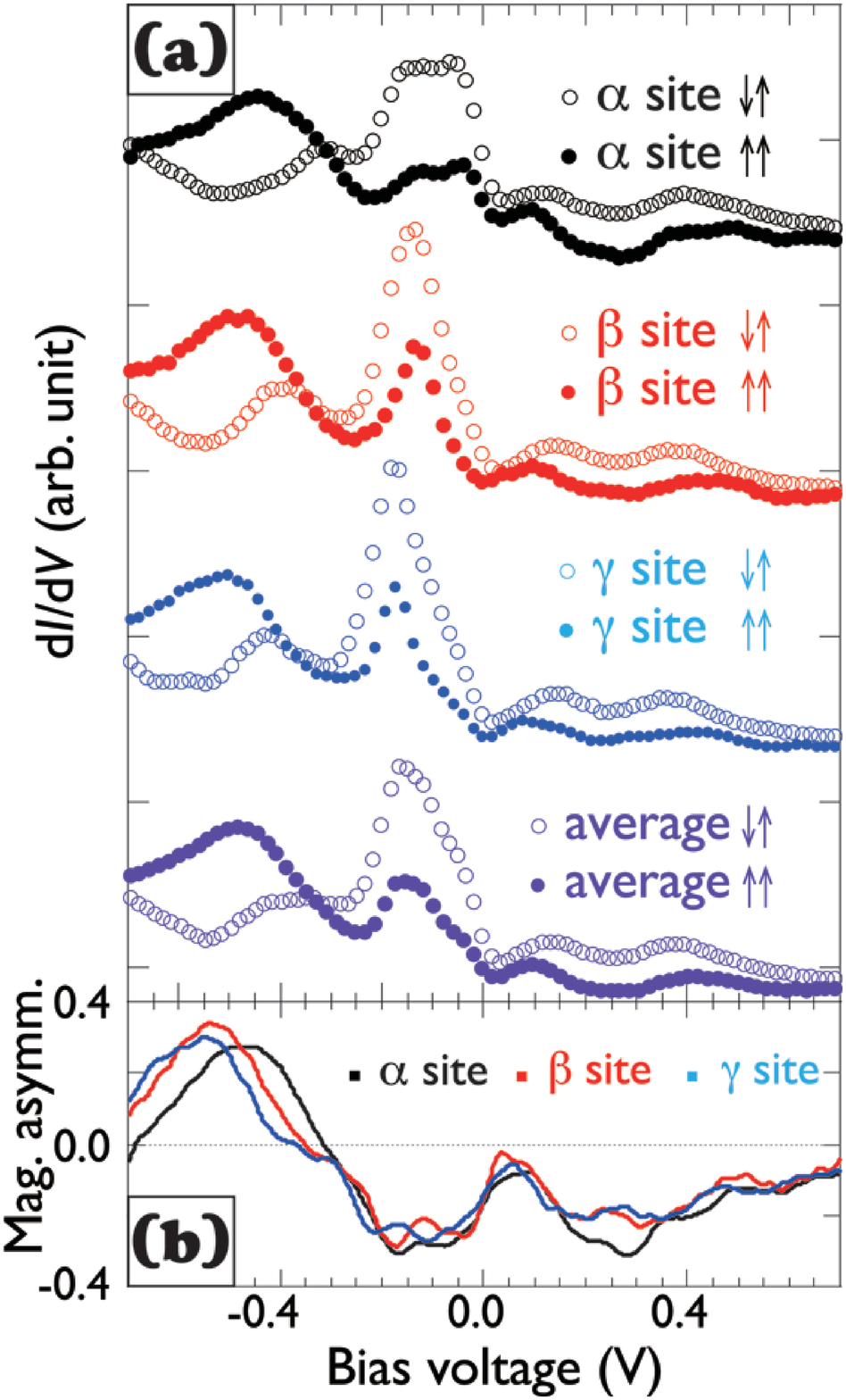}
\caption{\label{fig:fig5}
(Color online) 
(a) Spin-resolved d$I$/d$V$ spectra taken on three different sites of a strong moir\'e island and their averages.  
Arrows $\uparrow \uparrow$ ($\uparrow \downarrow$) refer to parallel (antiparallel) configurations of the tip and 
sample magnetizations. 
The spectra with the parallel and antiparallel configurations were obtained at 1T and 0.15 T (in the downward field sweep), respectively. 
The tip-sample separation is stabilized with $I_\mathrm{T}$= 1.0 nA and $V_\mathrm{S}$=-0.7 V. 
(b) Magnetic asymmetries arising from opposite magnetization configurations ($A=(\uparrow \uparrow- \uparrow \downarrow)/(\uparrow \uparrow + \uparrow \downarrow)$) at the three different sites. 
}
\end{center}
\end{figure}

To investigate their magnetic properties and responses to external magnetic fields, 
we utilized SPSTM technique with perpendicularly magnetized Cr bulk tip \cite{1st, annika, oka, romming}.  
First, we investigated strong-moir\'e islands by taking the d$I$/d$V$ mappings at -0.4 V, where the majority spin channel is located. 
A topographic STM image shown in the leftmost panel of Fig. 3 has both strong- and weak-moir\'e islands, which are marked with S and W, respectively. 
The thickness of the islands ranges from 6 to 8 ML. 
Three images on the right are d$I$/d$V$ mappings taken at the voltage of -0.4 V at the perpendicular magnetic fields of -0.5 T, -0.1 T, 
and -0.26 T (in the downward field sweep). 
One can easily notices changes in the d$I$/d$V$ contrast of strong-moir\'e islands marked with A to D in the three images. 
d$I$/d$V$ values of 4 strong-moir\'e islands in Fig. 3 in various magnetic fields are summarized in Fig. 4, 
together with that taken on the weak moir\'e island marked as E, which does not exhibit contrast variation. 
The absence of the spin contrast on the island E is simply because of no spin states at the bias voltage.  
It is not because of hydrogen contamination or segregation of substrate atoms onto the surface unlike the case of Co/Cu(111) \cite{h2, mt, rabe}, 
since the weak moir\'e islands also show the d$I$/d$V$ contrast variation at the different bias voltage 
as will be shown later in Fig. 6(a). 

The butterfly shape of the obtained curves for the islands A to D is a typical magnetoresistance curve in magnetic tunneling junctions \cite{MTJPRL, MTJJAP, MTJyuasa}, 
and was also reported in SPSTM studies \cite{ossie science, ouazi}.  
The shape is explained with the flippings of the magnetization directions of both the tip and the sample during the field sweep. 
Since all curves show abrupt changes at the magnetic field of -0.28 T and +0.25 T, 
the changes are attributed to flipping of the tip magnetization. 
Given the tip magnetization flipping, 
we can estimate curves of the islands themselves by removing the contribution of the bulk Cr tip from the butterfly curves. 
The inset of Fig. 4 is that of the island A demonstrating a typical shape of a ferromagnetic hysteresis curve. 
The other strong moir\'e islands also show the similar hysteresis curves with slightly different coercivities depending on the size and thickness. 
From these results, we conclude that the strong moir\'e islands 
have ferromagnetic properties with PMA. 

Spin-resolved d$I$/d$V$ spectra were also collected on a strong moir\'e island at fields 
where the tip and sample magnetizations are in parallel and antiparallel configurations as shown in Fig. 5(a) with solid and open circles, respectively.  
The spectra at all sites and all configurations exhibit peaks at the same energy positions with the spin-averaged spectra (Fig. 2(a)). 
In addition, the intensities vary between parallel and antiparallel configurations. 
The large peak close to the $E_\mathrm{F}$ is enhanced for the antiparallel configuration, 
while the small peak around -0.45 V for the parallel configuration \cite{ossie, berndt}. 
From the spectra, 
we estimated the magnetic asymmetry for each site (Fig. 5(b)). 
Overall trends are similar for all sites, 
however, 
clear shift in energy for the majority spin state is seen between $\alpha$ and the other sites, 
indicating that the spin-polarization of the Co island is spatially modulated by the moir\'e structure, 
similar with the case of the standing waves reported on Co/Cu(111) \cite{okascience}.

\begin{figure}[t]
\includegraphics[width=1\columnwidth,keepaspectratio]{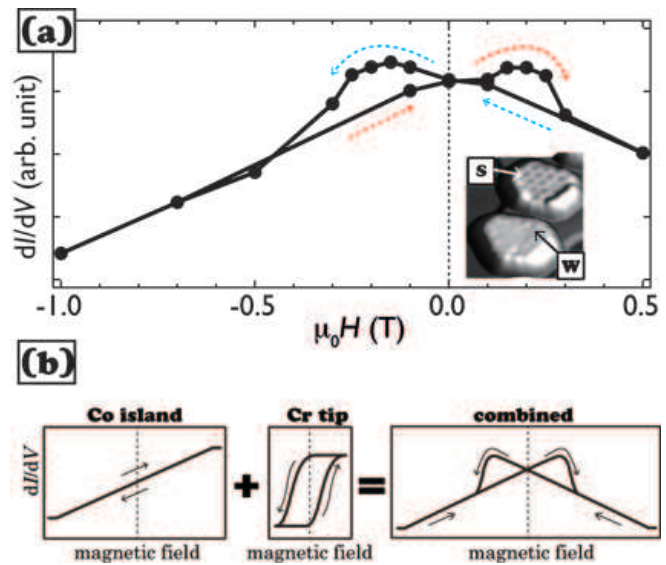}
\caption{\label{fig:in-plane mag}
(Color online) 
(a) Variation of the differential conductance d$I$/d$V$ at -0.2 V as a function of magnetic field taken on the weak moir\'e island (6 ML) shown in an STM image (inset, 19 nm x 22 nm, $V_\mathrm{S}$= -0.2 V, $I_\mathrm{T}$= 1 nA). 
The magnetic field was applied perpendicular to the surface plane. 
(b) Schematic diagrams to compose the shape of the curve in (a). The experimental data is qualitatively explained with combination of magnetization curves of in-plane-magnetized Co island and perpendicularly magnetized tip in out-of-plane magnetic fields.  
}
\end{figure}
Then, we focus on the magnetic property of the weak-moir\'e islands. 
The variation of the d$I$/d$V$ obtained from the spin-resolved d$I$/d$V$ mappings at -0.2 V on the weak moir\'e island (the inset of the Fig. 6(a)) 
at various magnetic fields is plotted in Fig. 6(a). 
The slope changes around $\pm$ 0.25 T most likely correspond to the 
flippings of the tip magnetization as observed in Fig. 4. 
The linear features up to -1.0 T suggest that the field was applied perpendicular to the easy axis of the magnetization direction and 
the magnetization was not saturated even beyond the saturation fields of the strong moir\'e islands ($\leq 0.3$ T). 
Therefore, the overall feature is understood by a combination of a linear magnetization curve of the Co island and a hysteresis one of the Cr bulk tip, 
as described schematically in Fig. 6(b). 
Since we applied magnetic fields perpendicular to the surface, this observation strongly suggests that the island is magnetized parallel to the surface. 
These observations reveal coexistence of perpendicularly and in-plane magnetized Co islands on the same surface.  
A recent surface-magneto-optical-Kerr-effect study also implied the possible coexistence of the in-plane and out-of-plane magnetizations on the same system but 
the spatial information was missing \cite{smoke}. 
The present results underline importance of local nanoscale structures on the overall magnetic properties and 
the powerful versatility of SPSTM for elucidating relevant issues.

\section{Theoretical analysis and discussions}

To understand the experimental observations, 
we performed theoretical calculations. 
Since it is known that fcc-stacked Co ultrathin films grown on a Cu(100) substrate shows ferromagnetic properties with the in-plane magnetic anisotropy \cite{kerkmann}, 
we speculated that the observed differences in the moir\'e corrugation amplitude and in the magnetization direction are due to difference in layer stackings of the Co islands. 
We performed the DFT calculations of three different stackings of of 5 ML Co thin film formed on 5 ML Ag(111), hcp (ABABA stacking), 
fcc (ABCAB stacking), and a pseudo-hcp (AB\textbf{C}BA stacking; hcp stacking with an fcc stacking fault), to see their structural stabilities. 
After the structural relaxation, 
we found out that the hcp structure is energetically most favorable and the fcc has the highest energy ($\Delta E=63 $ meV/atom). 
The pseudo-hcp structure was found stable enough ($\Delta E=39 $ meV/atom) to be formed 
in the case of the room temperature deposition. 

\begin{figure}[b]
\includegraphics[width=1.0\columnwidth,keepaspectratio]{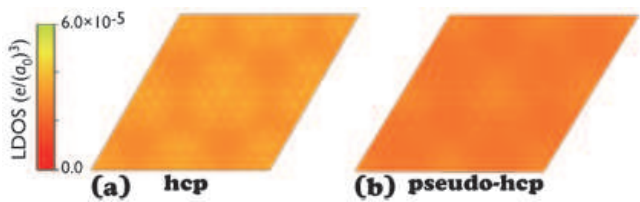}
\caption{\label{fig:fig7}
(Color online) 
Local density of states integrated between -0.2 eV and $E_\mathrm{F}$ 
in a plane parallel to and $\sim$ 0.27 nm above the surface,  
for (a) hcp with ABABA stacking and (b) pseudo-hcp with AB\textbf{C}BA stacking. 
$a_\mathrm{0}$ is the Bohr radius. 2 $\times$ 2 supercells of the moir\'e pattern are shown. 
}
\end{figure}

\begin{figure}[htbp]
\includegraphics[width=0.85\columnwidth,keepaspectratio]{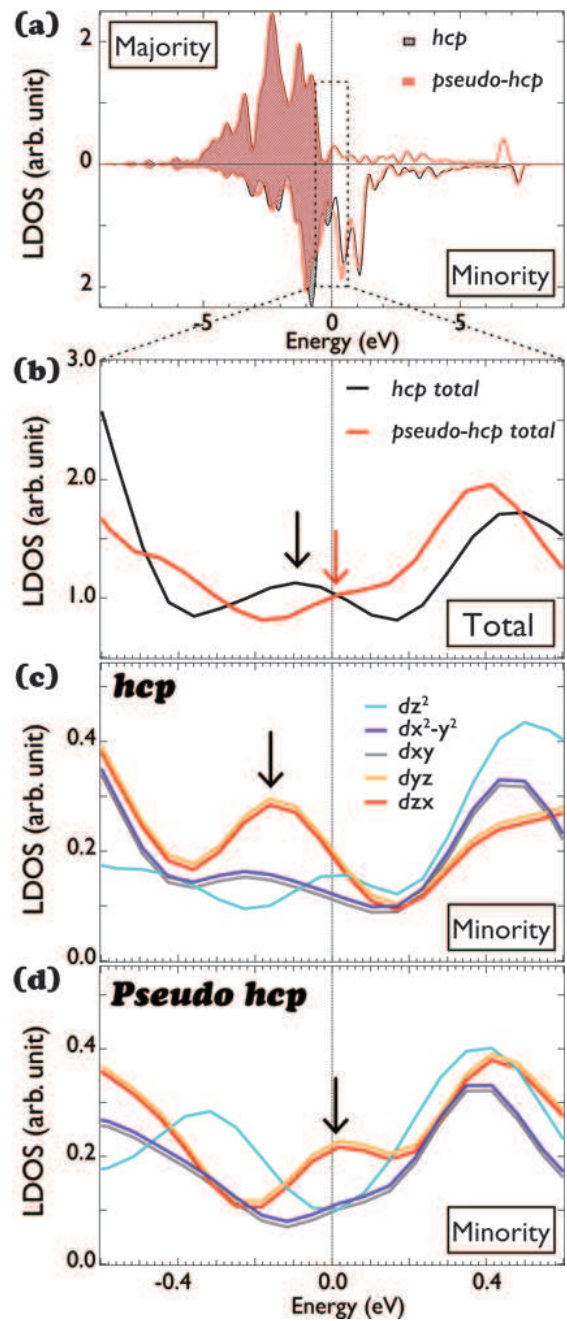}
\caption{\label{fig: cal LDOS} 
(Color online) 
(a) Calculated LDOS of the Co atom at the $\alpha$ sites of 5 ML Co on 5 ML Ag(111) with hcp and pseudo-hcp stackings. 
(b) Calculated LDOS around the $E_\mathrm{F}$. 
The total LDOSs including those of the majority- and minority-spin bands are shown. 
The arrows indicate the peak and shoulder around the $E_\mathrm{F}$ which are observed in the experiment (Fig. 2).
(c, d) Decompositions of the LDOS of the minority-spin bands into $d$ orbitals. (c) and (d) are for hcp and pseudo-hcp stackings, respectively. 
The arrows indicate the energy of the peak and shoulder around the $E_\mathrm{F}$ in (b).  
In (c) and (d), the contributions from {\it s} and {\it p} orbitals are not shown since they are negligible compared to those of {\it d} orbitals. 
We show only the minority-spin band since there is not much difference between the two stackings for the majority-spin band.  
The $d_\mathrm{xy}$ and $d_\mathrm{x^{2}-y^{2}}$ states, $d_\mathrm{yz}$ and $d_\mathrm{zx}$ states are degenerated but 
they are plotted with small shifts for the clarity. }
\end{figure} 

Next, we calculated integrated LDOS of the hcp and pseudo-hcp structures as shown in Fig. 7 
in the energy range between $E_\mathrm{F}$ and -0.2 eV in the plane parallel to and $\sim$ 0.27 nm above the surface. 
2 $\times$ 2 supercell of the moir\'e pattern are shown here. 
The hcp structure clearly shows the stronger moir\'e contrast than the pseudo-hcp. 
The calculations indicate that the stacking fault in hcp Co islands modifies the electronic structure, and as a result, the moir\'e corrugation amplitude is weakened. 

The LDOS at the Co atom at the $\alpha$ site was also calculated for hcp and pseudo-hcp stacked 5 ML Co layers on 5 ML Ag(111) as shown in Fig. 8, 
to investigate the origin of the difference in the moir\'e corrugation amplitude between the two stackings seen in Fig. 7. 
While no clear difference is seen between the majority-spin bands of the two stackings, clear shifts of the states around and below the $E_\mathrm{F}$ can be recognized in the minority-spin bands (Fig. 8(a)). 
In the total LDOS including the majority- and minority-spin band contributions around the $E_\mathrm{F}$ (Fig. 8(b)), 
which directly corresponds to the experimental d$I$/d$V$ spectra, 
a distinct peak can be identified around -0.2 eV for the hcp case as we observed experimentally for the strong moir\'e island (Fig. 2(a)). 
The peak changes to the shoulder and shifts to the $E_\mathrm{F}$ for the pseudo-hcp case as the case of the weak moir\'e island in our experiment (Fig. 2(b)). 
By projecting the LDOS to the atomic orbitals, 
we can understand the origin of the peak shift between strong and weak moir\'e islands (Figs. 8(c, d)). 
Around -0.2 eV, $d_\mathrm{yz}$ and $d_\mathrm{zx}$ orbitals have peaks for the hcp case and shifts to the $E_\mathrm{F}$ for the pseudo-hcp case. 
Therefore, we conclude that the peak shift between strong and weak moir\'e islands is caused by 
the shifts in minority-spin states due to the $d_\mathrm{yz}$ and $d_\mathrm{zx}$ orbitals. 

Finally, we discuss the magnetic anisotropy in order to understand the difference in magnetic properties between the strong and weak moir\'e islands. 
Here we consider two types of magnetic anisotropies; one is magnetocrystalline 
anisotropy due to the spin-orbit interaction, and the other is shape anisotropy due to the magnetic dipole interaction. 
The magnetocrystalline anisotropy was estimated by DFT calculations 
for the two stacking structures with free-standing 5 atomic layers of Co. 
While the hcp structure has PMA ($\Delta E_\mathrm{hcp}^\mathrm{crystal}=0.084$ meV/Co atom), 
it is reduced by a factor of more than 2 in the pseudo-hcp structure ($\Delta E_\mathrm{pseudo-hcp}^\mathrm{crystal}=0.034$ meV/Co atom). 
This result is reasonable in comparison with an experimental report on a Co-based alloy in which 10 \% of fcc stacking faults reduces the PMA by about factor 2 revealed 
by the in-plane X-ray diffraction and magnetization measurements \cite{sf hinata}. 
The shape anisotropy, which favors in-plane magnetization for the island dimensions in this study, was estimated in the framework of a rotationally symmetric ellipsoid model 
with a calculated magnetic moment $1.64 \mu_\mathrm{B}$/Co atom ($1.70 \mu_\mathrm{B}$/Co atom) for hcp (pseudo-hcp) structure 
with free-standing 5 atomic layers of Co \cite{cal moment}. 
It turns out that the estimated shape anisotropy ($\Delta E_\mathrm{hcp}^\mathrm{shape} \sim 0.067$ meV/Co atom and $\Delta E_\mathrm{pseudo-hcp}^\mathrm{shape} \sim 0.071$ meV/Co atom) is smaller than magnetocrystalline anisotropy of the hcp structure but bigger than that of the pseudo-hcp. 
This indicates that in-plane magnetization is favorable for the pseudo-hcp, while perpendicular magnetization for the hcp. 

Based on these discussions, 
we conclude that the effect of fcc stacking fault in hcp Co nanoscale islands are 
significant from both the electronic and magnetic points of view; the modification of the electronic structure reduces the corrugation amplitude of the moir\'e pattern, 
and the drastic reduction in the PMA induces the in-plane magnetized Co islands coexisting with hcp and perpendicularly magnetized ones. 
\section{Conclusions}
In summary, 
electronic and magnetic influences of an fcc stacking fault in hcp Co nanoscale islands on the Ag(111) surface are addressed 
by utilizing SPSTM experiments and DFT calculations. 
We have observed Co nanoscale islands with stronger and weaker moir\'e corrugation amplitudes at $V_\mathrm{S}=-0.2$ V. 
The spin-averaged d$I$/d$V$ spectra indicate that the islands have the different electronic structures depending on the moir\'e corrugation amplitudes. 
The spin-polarized d$I$/d$V$ images of the islands and the magnetic-field dependence 
clarify that the islands with the stronger moir\'e corrugation amplitude are perpendicularly magnetized 
and the ones with the weaker moir\'e corrugation amplitude are in-plane.  
DFT calculations reproduce the two types of islands with respect to the moir\'e corrugation amplitude and the magnetic anisotropy 
by considering two different stacking structures for the islands; hcp stacking and that with an fcc stacking fault.  
It strongly suggests that the effect of an fcc stacking fault in an intrinsic hcp stacking are significant from both the electronic and magnetic points of view. 
These results give important microscopic information to design and construct the future ultrahigh density magnetic recording devices made with Co or Co-based alloys. 

\section*{acknowledgement}
	
We thank Christian Henneken for technical information about a Cr bulk tip. We also thank Kirsten von Bergmann, Susumu Shiraki, Howon Kim, Stefan Bl\"ugel and Yoshio Miura for fruitful discussions. 
This work is partially funded by Grants-in-Aid for Scientific Research, Japan Society for the Promotion of Science (21360018, 23840008, 25110008, 25286055, 25707025, 25871114, 26120508, 26110507) 
and Foundation Advanced Technology Institute.  
The calculations were performed with the computer facilities of the Institute of Solid State Physics (ISSP Super Computer Center, University of Tokyo), Numerical Materials Simulator at NIMS, and RIKEN Integrated Cluster of Clusters (RICC).


\begin{thebibliography}{44}
\bibitem{cal moment}  The calculated magnetic moment for the pseudo-hcp is enhanced by $\sim$4 \% in comparison to that of the hcp. 
Similar enhancement of the magnetic moment in the fcc structure was reported in the DFT calculation for the bulk Co crystal \cite{PRL moment}.
\end{thebibliography}
\end{document}